# Probing magnetic symmetry in antiferromagnetic $Fe_4Nb_2O_9$ single crystals by linear magnetoelectric tensor

Jing Zhang,[1] Na Su,[2] Xinrun Mi, Maocai Pi,[1] Haidong Zhou,[3] Jinguang Cheng[2], and Yisheng Chai[1*]

[1]*Low Temperature Physics Laboratory, College of Physics, Chongqing University, Chongqing 401331, China*

[2]*Beijing National Laboratory for Condensed Matter Physics and Institute of Physics, Chinese Academy of Sciences, Beijing 100190, China.*

[3]*Department of Physics and Astronomy, University of Tennessee, Knoxville, Tennessee 37996, USA*

[*]yschai@cqu.edu.cn

## Abstract

In the present study, we investigated magnetodielectric, magnetoelectric (ME) and angular dependent polarization in single crystal $Fe_4Nb_2O_9$. The magnetodielectric effects in $\varepsilon_x$ ($x$//[100]), $\varepsilon_y$ ($y$//[120]) and $\varepsilon_z$ ($z$//[001]) are found to be significant only around $T_N \approx 95$ K when magnetic fields are applied along three orthogonal $x$, $y$ ($y$//[120]) and $z$ directions. The finite polarization $P_x$, $P_y$ and $P_z$ of 70 μC/m$^2$, 100 μC/m$^2$ and 30 μC/m$^2$ respectively, can be induced in the antiferromagnetic phase when a finite magnetic field up to 9 T was applied along the three orthogonal directions. The angular dependent polarization measurements verify the dominating linear ME effects below $T_N$. From the above experimental results, a linear ME tensor $a_{ij}$ with all nine non-zero components can be inferred, demonstrating a much lower magnetic point group of **-1'** for the canted antiferromagnetic configuration.

# I. INTRODUCTION

The direct linear magnetoelectric (ME) effect, where the electric polarization $P$ is induced in proportion by an external magnetic field $H$ [1,2], is expressed as $P_i = \alpha_{ij}H_j$ ($\alpha_{ij}$ is linear ME tensor where $i$ and $j$ run over all the Cartesian coordinates, $x$, $y$, $z$) [3,4]. It has attracted great interests due to the potential application for low-power-consumption sensor devices and novel physical properties in solids [5-8] [6,9-12]. This effect can exist in solids with long-range magnetic order, which breaks the space inversion and time reversal symmetry simultaneously, like the time-honored $Cr_2O_3$[13] with antiferromagnetic (AFM) order. Below $T_N$ = 307 K, $Cr_2O_3$ belongs to the magnetic point group **-3'm'**. From the Newmann's principle, the nonzero ME tensor components are $\alpha_{xx}$, $\alpha_{yy}$ and $\alpha_{zz}$. Only 58 out of 90 magnetic point groups allow the non-zero $\alpha_{ij}$ [14]. In this regard, symmetry analysis contributes to the searching for the linear ME effects in magnetic materials [1].

Recently, the investigation of $A_4Nb_2O_9$ (A= Mn, Co, Fe) has drawn strong attentions as a potential ME material system, which belongs to the corundum-type structure firstly found by Bertaut *et al.* [15,16]. They crystallize in a trigonal crystal structure with the space group *P-3c1*, a derivative of the α-$Al_2O_3$-type structure [15,17]. Two kind of A atoms, A1 and A2 are located at 4d sites while Nb and two O atoms at 4c, 6g and 12f Wyckoff positions, respectively, as shown in Fig. 1a. The magnetic structures of both Mn and Co cases in powered samples were first determined to be a collinear AFM configuration along the *c*-axis [15]. Their magnetic point group is **-3'm'** too, and have been confirmed in experiments [2,5-7].

However, $Fe_4Nb_2O_9$ shows an in-plane AFM configurations. The magnetic cation $Fe^{2+}$ has a $3d^6$ configuration with $S = 2$ and is found to be ordered at $T_N \approx$ 90 K (94 K, 97 K) by magnetic susceptibility [18-20]. It experiences another structure phase transition around 77 K (80 K, 70 K), as manifested by an anomaly in dielectric constant and verified by Powder neutron diffraction experiment [19]. Below 77 K, its space group is lowered to ***C2/c*** (point group of **2/m**), which is not presented in the Co and Mn cases. In Ref. [19], the magnetic structure just below $T_N$ is described by collinearly arranged moments in the *ab* plane parallel to either *x*//[100] or *y*//[120] directions, which can be assigned in either **2/m'** or **2'/m** magnetic point group, as shown in Figs. 1b and 1c, respectively. The matrix forms of the ME tensor for **2/m'** and **2'/m** should be [19]:

$$\alpha_{ij} = \begin{bmatrix} \alpha_{xx} & 0 & \alpha_{xz} \\ 0 & \alpha_{yy} & 0 \\ \alpha_{zx} & 0 & \alpha_{zz} \end{bmatrix} \text{ and } \alpha_{ij} = \begin{bmatrix} 0 & \alpha_{xy} & 0 \\ \alpha_{yx} & 0 & \alpha_{yz} \\ 0 & \alpha_{zy} & 0 \end{bmatrix}, \text{ respectively} \quad (1)$$

Later on, a single-crystal neutron diffraction technique with refined magnetic structure below $T_N$ reveals that the adjacent Fe1-Fe2 spin pairs along $z$ direction canted a 5.81° in the plane, as shown in Figs.1d and 1e. The magnetic space group has been described as ***C2/c'*** (magnetic point group of ***2/m'***)[20]. Indeed, both polycrystalline and single-crystal samples were found to show magnetic-field induced polarization below $T_N$, confirming the existence of at least one of the non-zero ME coefficients in this compound. Besides, Ref. [20] also indicates that $Fe_4Nb_2O_9$ is a linear ME material by soft phonon mode. However, to pin down the exact magnetic structure and related point group, one has to measure all nine $\alpha_{ij}$ components of this compound.

In this letter, we investigate the dielectric, pyroelectric and angle-dependent polarization under different external $H$ in two $Fe_4Nb_2O_9$ single crystals. We found that $Fe_4Nb_2O_9$ exhibits (i) anisotropic magnetodielectric effects, consistent with Ref. [20], (ii) nine nonzero components in linear ME tensor $\alpha_{ij}$ reflected by $H$ induced polarization below 95 K and (iii) polarization with sinusoidal oscillation induced by rotating magnetic field below $T_N$ for $\mu_0 H$ = 6 T. All these results show that $Fe_4Nb_2O_9$ is a linear ME material with magnetic point group -**1'** by ymmetry analysis.

## II. METHODS

The single crystals of $Fe_4Nb_2O_9$ were grown from the traveling solvent floating-zone technique [20]. For electrical measurements, two single-crystal were cut into a rectangular shape of 0.6×1.4×1.08 $mm^3$ and 0.9×2.52×0.42 $mm^3$, with the largest surface normal to the $x$//[100] and $z$//[001] directions, respectively. Gold electrodes were deposited onto these faces for electric measurements. Relative dielectric constant (ε) measurements were performed at 100 kHz using an Agilent E4980A Precision LCR meter. Before the pyroelectric current ($I$) measurements, the specimen went through the ME annealing procedure from 99 K. For +$H_p$ & ±$E$ annealing condition, the sample was kept electrically poled with +$\mu_0 H_p$ = 0 T, 1 T, 2 T, 4 T, 9 T, ±$E_1$ = ±183 kV/m (//$x$ or //$z$) and ±$E_2$ = ±79.4 kV/m (//$y$, due to much longer geometry along this direction) with external $H$ down to 77 K. Then, ±$E_1$ or ±$E_2$ was turned off and the measurements were performed by sweeping $T$ to cross the boundary between ferroelectric and paraelectric region, *i.e.* from 77 K to 99 K on warming at a rate of 1 K/min using a Keithley 6517B electrometer. The temperature-dependent pyroelectric currents, $I_x$, $I_y$ and $I_z$ along $x$, $y$ and $z$ directions respectively, were measured and integrated as a function of time to determine the $P$ under selected $H$ values and orientations. The linear ME tensor components are estimated from average 9 T data in each pair of $+H_p$ & $\pm E$ configurations. The temperature and the magnetic field were controlled using a Dynacool system with 9 T superconducting magnet (Quantum Design). Angular dependent polarization measurements were performed on a rotator (MultiField Tech.) in Dynacool system.

## III. RESULTS AND DISCUSSIONS
### A. Dielectric constant measurements

In previous works, it is revealed that $Fe_4Nb_2O_9$ undergoes an antiferromagnetic and a structure phase transition at $T_N$ = 90 K (94 K, 97 K) and 77 K (80 K, 70 K), respectively [18-20]. In order to investigate the magnetoelectric properties in the AFM phase below $T_N$, we investigate the $T$-dependent $\varepsilon_i$ under selected $H_j$ ($i, j = x, y$ and $z$) in two $Fe_4Nb_2O_9$ single crystals, as shown in Fig. 2. For $\varepsilon_x$ in warming, it quickly increases below 88 K and becomes flat until $T_N$ = 95 K where a small kink was observed for $\mu_0H$ = 0 T. This is consistent with the reported behavior in Ref. [20] with 20 kHz. When an external magnetic field is applied along the $x$ direction, no change of $\varepsilon_x$ can be observed except a small temperature region around $T_N$ (92 to 96 K). Up to 9 T, the small kink around $T_N$ becomes a clear bump, as indicated in the inset of Fig. 2a. Our data has a better resolution than that in Ref. [20] in which a magnetodielectric effect with the same configuration of $E//H//x$ cannot be observed clearly. When an external magnetic field up to 9 T is applied along the $y$ direction, the change of $\varepsilon_x$ is almost negligible in the entire measured $T$ region. However, for $H//z$ configuration, weaker bump features around $T_N$ can be discerned as shown in the inset of Fig. 2c. No strong magnetodielectric effects can be discerned in other temperature regions. For $T$ dependent $\varepsilon_y$ with $H//x, y$ and $z$, they are almost identical with that of $\varepsilon_x$. It is due to the very weak in-plane crystal anisotropy in hexagonal lattice.

For $T$ dependent $\varepsilon_z$ with $\mu_0H$ = 0 T, $\varepsilon_z$ increases monotonically from 75 K to 99 K while no anomaly can be found around $T_N$, consistent with the previous report [20]. When an external magnetic field along $x$ direction is applied up to 9 T, a clear peak in $\varepsilon_z$ is induced around $T_N$, as shown in Fig. 2g while $\varepsilon_z$ in the entire $T$ region is slightly enhanced with a small offset. For $H//y$ configuration, the magnetodielectric behavior is very similar to that of $H//x$ configuration. Note that the $\mu_0H$ = 0 T data are slightly different among three $H$ configurations due to the reloading of the sample at room $T$. For $H//z$ configuration, the magnetodielectric effect is weakest with a small bump around $T_N$ up to 9 T. We can conclude that the clear magnetodielectric effects of $Fe_4Nb_2O_9$ around $T_N$ strongly indicate the existence of ME effects in all nine configurations below 95 K, which is inconsistent with our previous study [19]. Note that the small feature around 88 K may indicate some non-magnetic phase transition which requires further investigations.

### B. Magnetoelectric Effects

To investigate the ME effects in the AFM phase just below $T_N$ in this compound, we performed a comprehensive $T$-dependent polarization measurement along $P//x$, $y$ and $z$ axis under different $H$ orientations, as shown in Figure 3. From the above configurations, we expect to probe nine components in the linear ME tensor $\alpha_{ij}$, where **2'/m** and **2/m'** magnetic point groups allow conjugate non-zero components in Eq. 1. Without bias $H$, no ferroelectricity can be observed along crystal $x$, $y$ and $z$ directions. Under finite $\mu_0 H$ up to 9 T, $P_x$, $P_y$ and $P_z$ can be gradually induced up to about 70 μC/m$^2$, 100 μC/m$^2$ and 30 μC/m$^2$ respectively, without saturation. Moreover, the induced $P$ values can be fully reversed by reversing the poling $E$, consistent with reference [20], which indicates that $Fe_4Nb_2O_9$ is a magnetoelectrics in the AFM phase. Surprisingly, even though the magnetodielectric effects are negligible for $H//z$ configuration for $\varepsilon_y$ and $\varepsilon_z$, the induced $P_y$ and $P_z$ values are comparable or even larger than that in $H//x$ and $y$ directions for 9 T data, which requires further investigations. $P_x$, $P_y$ and $P_z$ values at 90 K show quasi-linear dependence with magnetic field along all three orthogonal directions, as shown in the insets of Fig. 3, implying dominating linear ME effects. To further verify this, we performed a reversal of polarization by reversing magnetic field experiments below $T_N$, as shown Fig. 4e. The sample is first ME poled by $+E$ and $\mu_0 H = +9$ T down to 77 K. Then it will be warmed up to 99 K and 9 T with measuring the pyroelectric currents $I_x$ and $I_z$. Or, the magnetic field will sweep to -9 T at 77 K, and the sample is warmed up to 99 K and -9 T with measuring the pyroelectric currents. As shown in Figs. 4a and c, the sign of $I_x$ and $I_z$ are reversed by reversing the $\mu_0 H//x$ and $\mu_0 H//z$ from 9 T to -9 T. Accordingly, the $P_x$ and $P_z$ can be fully reversed, as shown in Fig. 4b and 4d, respectively. The polarization reversal by magnetic field again points to the dominating linear or odd-order ME effects in this system and consistent with the similar previous report with $P_z$ and $\mu_0 H//x$ configuration [20]. We will show below that the linear ME effects are dominating in this system. From the 9 T and 90 K data in Fig. 3, the linear tensor components of $\alpha_{xx}$, $\alpha_{xy}$, $\alpha_{xz}$, $\alpha_{yx}$, $\alpha_{yy}$, $\alpha_{yz}$, $\alpha_{zx}$, $\alpha_{zy}$, $\alpha_{zz}$ are estimated to be 8.9, 13.0, 9.0, 13.4, 13.6, 15.1, 4.5, 4.5, 4.4 ps/m respectively in absolute values, pointing to a much lower magnetic point group at 90 K. We have to point out that the $P$ and related ME tensor components are almost fully saturated under poling *E* fields applied.

### C. Angular Dependent Polarization

To unambiguously distinguish between the linear and higher odd order ME effects, we perform an angular (*θ*) dependence of polarization $P_x$ under rotating *H* experiments in *x-z* plane as an example. Linear, third and higher-order terms will result in a function of sin*θ*, sin3*θ* and sin5*θ*…terms in $P_x$, respectively. Particularly, for the linear ME tensor, $P_x = \alpha_{xx}H_x + \alpha_{xz}H_z$. Here, the magnetic field $\mu_0 H$ = 6 T rotates from *x* to *z* directions, as shown in Figs. 4f and g. We can clearly see that the angular dependent $P_x$ shows a dominating sinusoidal behaviors with small hysteresis between *θ* increase and decrease runs. This is a direct proof of dominating linear ME effect in this system. In this configuration, $H_x = H\cos\theta$ and $H_z = H\sin\theta$, leading to:

$$P_x = H(\alpha_{xx}\cos\theta + \alpha_{xz}\sin\theta) = H\sqrt{(\alpha_{xx}^2 + \alpha_{xz}^2)}\cos(\theta - \theta_0) \qquad (2)$$

where $\theta_0 = \arctg(\alpha_{xz}/\alpha_{xx})$. From Fig. 4g, the small average offset value of $\theta_0 \approx -32°$ can be deduced from the angle increase and decrease runs, indicating an $\frac{\alpha_{xx}}{\alpha_{xz}} \approx -1.6$ ratio value and being consistent with the calculated values of $\left|\frac{\alpha_{xx}}{\alpha_{xz}}\right| = 8.9/9.0 \approx 1.0$ from the data in Fig. 3. The negative sign of $\theta_0$ reveals that the $\alpha_{xx}$, and $\alpha_{xz}$ has the opposite sign in same AFM domain. The hysteresis behaviors may be related to the magnetic anisotropy between in-plane and out-of-plane.

### D. Symmetry Analysis

From the above results, $Fe_4Nb_2O_9$ is an intriguing linear ME material with all nine non-zero terms, which is in direct contrast to the reported magnetic point group of **2'/m** or **2/m'** which has at most 4 and 5 non-zero components, respectively. Recently, a non-collinear magnetic configuration with a 5.81° canting between the adjacent Fe1-Fe2 spin pairs along *z* direction was proposed through single-crystal neutron diffraction, as shown in Fig. 1e. Its magnetic point group was described as **2/m'**, while a more careful reexamination in this paper indicates a **-1'** magnetic point group. In this configuration, space inversion centre+time reversal in Fe1 layers are retained regardless of the spins tilting in this layer. However, the 2*y* operation in the middle of Fe2 layers is broken due to canting between Fe2 spins. The overall magnetic point group of $Fe_4Nb_2O_9$ is **-1'** with a linear ME matrix form:

$$\begin{bmatrix} \alpha_{xx} & \alpha_{xy} & \alpha_{xz} \\ \alpha_{yx} & \alpha_{yy} & \alpha_{yz} \\ \alpha_{zx} & \alpha_{zy} & \alpha_{zz} \end{bmatrix} \qquad (3)$$

where all nine coefficients can be non-zero. This matrix form is consistent with our experimental results of nine non-zero components. There is only two magnetic point group **1**, and **-1'** can allow all nine ME components while the magnetic order at least preserves -1' symmetry. In this sense, we are confident to conclude a **-1'** magnetic point group at 90 K.

## IV. CONCLUSION

In summary, we studied the magnetodielectric and magnetoelectric properties of $Fe_4Nb_2O_9$. The relative dielectric constants along $x$//[100], $y$//[120] and $z$//[001] show strong field dependent behaviors around $T_N$. Accordingly, all nine linear ME tensor components are estimated to have finite values. From symmetry analysis, the magnetic point group of the antiferromagnetic configuration below $T_N$ is determined to be **-1'** instead of **2/m'**. Our work suggests that measurements of ME tensor is a practical approach to assist the determination of long-range magnetic order in the insulating magnetic materials.

## ACKNOWLEDGMENTS

This work is supported by the Natural Science Foundation of China grant Nos. 11674384, 11974065, 11904040, 12004056, 12025408, 11921004, 11874400. This work has been supported by Chongqing Research Program of Basic Research and Frontier Technology, China (Grant No. cstc2020jcyj-msxmX0263), Fundamental Research Funds for the Central Universities, China (2020CDJQY-A056, 2020CDJ-LHZZ-010, 2020CDJQY-Z006), Projects of President Foundation of Chongqing University, China(2019CDXZWL002). We Would like to thank Miss G. W. Wang at Analytical and Testing Center of Chongqing University for her assistance. H.D.Z. acknowledges support from NSF-DMR-2003117.

# REFERENCES

[1] H. Schmid, *Ferroelectrics* **162**, 317 (1994).

[2] J.-P. Rivera, Eur. Phys. J. B **71**, 299–313 (2009)

[3] Y. Tokura, S.Seki, and N. Nagaosa, Rep. Prog. Phys. **77** (2014) 076501

[4] T. H. O'Dell, *The Electrodynamics of Magneto-Electric Media* (North-Holland, Amsterdam, 1970).

[5] M. Fiebig, J. Phys. D **38**, R123 (2005).

[6] Y. Tokura, S. Seki, and N. Nagaosa, Rep. Prog. Phys. **77**, 076501 (2014).

[7] N. A. Spaldin, M. Fiebig, and M. Mostovoy, J. Phys.: Condens. Matter **20**, 434203 (2008).

[8] H. Schmid, J. Phys.: Condens. Matter **20**, 434201 (2008).

[9] J. Wang, J. B. Neaton, H. Zheng, V. Nagarajan, S. B. Ogale, B. Liu, D. Viehland, V. Vaithyanathan, D. G. Schlom, U. V. Waghmare, N. A. Spaldin, K. M. Rabe, M. Wuttig, R. Ramesh, Science **299**, 1719 (2003).

[10] N. A. Spaldin, M. Fiebig, Science **309**, 391 (2005).

[11] W. Eerenstein, N. D. Mathur, and J. F. Scott, Nature (London) **442**, 759 (2006).

[12] S.-W. Cheong and M. Mostovoy, Nat. Mater. **6**, 13 (2007).

[13] T. R. McGuire, E. J. Scott, and F. H. Grannis, Phys. Rev. **102**, 1000 (1956).

[14] R. E. Newnham, *Properties of Materials: Anisotropy, Symmetry, Structure* (Oxford University Press, USA, 2005).

[15] E. F. Bertaut, L. Corliss, F. Forrat, R. Aleonard, and R.Pauthenet, J. Phys. Chem. Solids **21**, 234 (1961).

[16] E. Fischer, G. Gorodetsky, and R. M. Hornreich, Solid State Commun. **10**, 1127 (1972).

[17] N. D. Khanh, N. Abe, H. Sagayama, A. Nakao, T. Hanashima, R. Kiyanagi, Y. Tokunaga, and T. Arima, Phys. Rev. B **93**,075117 (2016).

[18] A. Maignan and C. Martin, Phys. Rev. B **97**, 161106(R) (2018).

[19] R. Jana, D. Sheptyakov, X. Y. Ma, J. A. Alonso, M. C. Pi, A. Muñoz, Z. Y. Liu, L. L. Zhao, N. Su, S. F. Jin, X. B. Ma, K. Sun, D. F. Chen, S. Dong, Y. S. Chai, S. L. Li, and J. G. Cheng, Phys. Rev. B **100**, 094109 (2019).

[20] L. Ding, M. Lee, E. S. Choi, J. Zhang, Y. Wu, R. Sinclair, B. C. Chakoumakos, Y. S. Chai, H. D. Zhou, and H. B. Cao, Phys. Rev. Mater. **4**, 084403 (2020).

**FIGURE CAPTIONS**

**FIG. 1.** (a) Schematic crystal structure of $Fe_4Nb_2O_9$. (b,c,d) Top view of magnetic structures with spin along $y$, $x$ and canted directions, respectively. (e) Side view of spin configuration in (d) with space inversions centers (marked as X) in Fe1 layers.

**FIG. 2.** The temperature dependent relative dielectric constant of $Fe_4Nb_2O_9$ $\varepsilon_x$ for (a) $H//x$, (b) $H//y$ and (c) $H//z$, $\varepsilon_y$ for (d) $H//x$, (e) $H//y$ and (f) $H//z$, and $\varepsilon_z$ for (g) $H//x$, (h) $H//y$ and (i) $H//z$. All the data are measured in warming. Insets of (a)-(f) and (i) show the data around 94 K.

**FIG. 3.** The temperature dependent $P_x$, $P_y$ and $P_z$ under orthogonal $H$ along the (a), (d), (g) $x$ direction, (b), (e), (h) $y$ direction and (c), (f), (i) $z$ direction, respectively. Three insets plot the $H$ dependent polarization values at 90 K calculated from (a)-(i).

**FIG. 4.** The temperature dependent pyroelectric currents $I_x$ (a) and $I_z$ (c) in $H//x$ and $H//z$, respectively. The temperature dependent polarizations $P_x$ (b) and $P_z$ (d) in $H//x$ and $H//z$, respectively. (e) Schematic poling process of $H$ reversing from 9 T to negative 9 T. (f) Angular dependence of polarization under the rotating $\mu_0 H = 6$ T at $T = 90$ K. Schematic sample configuration is shown in the upper panel.

**FIGURES**

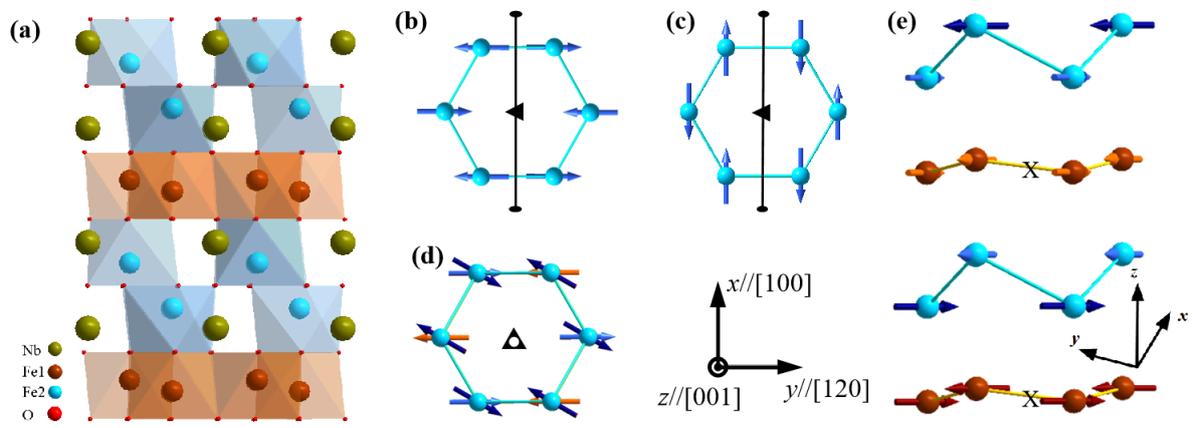

Figure 1 Jing Zhang *et al*.

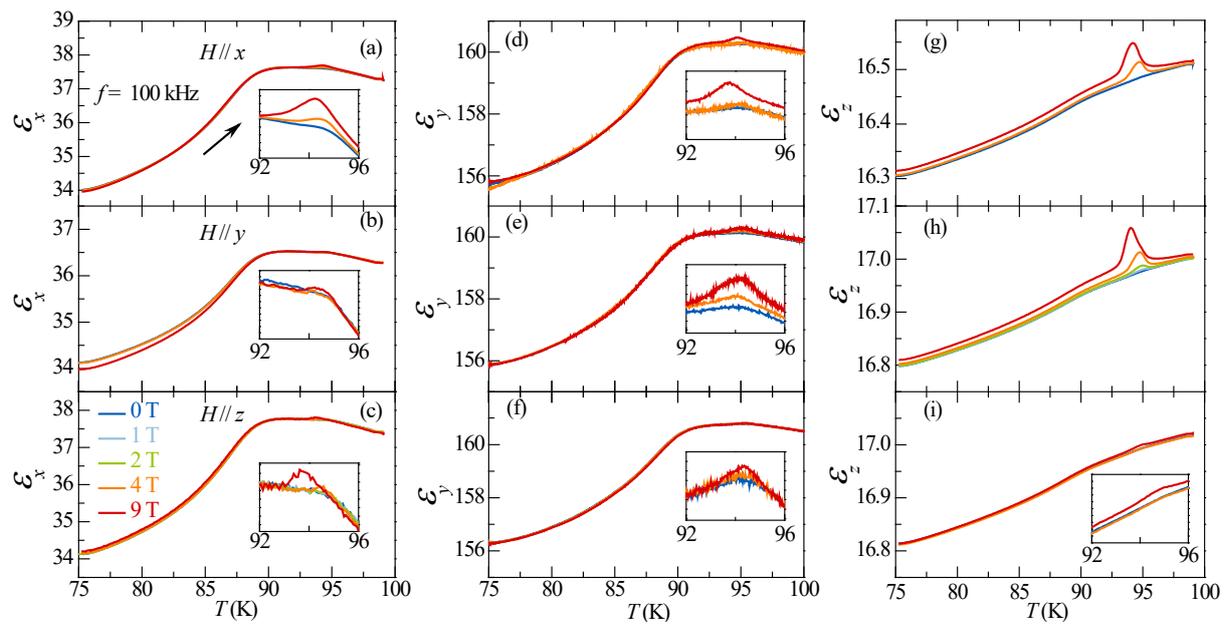

Figure 2 Jing Zhang *et al.*

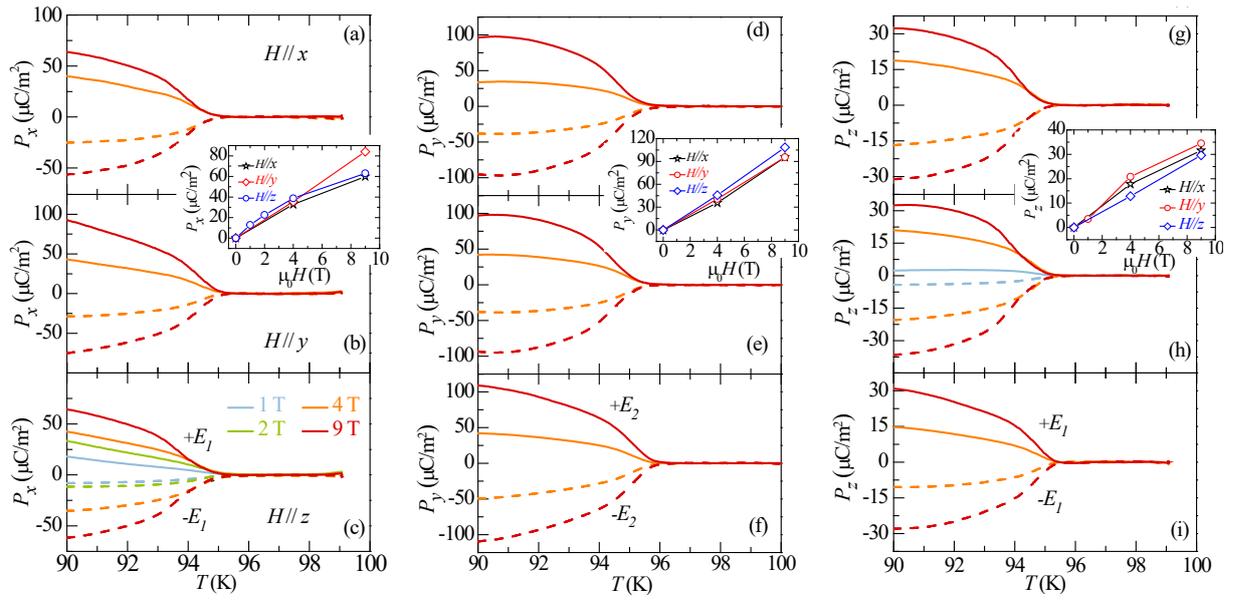

Figure 3 Jing Zhang *et al*.

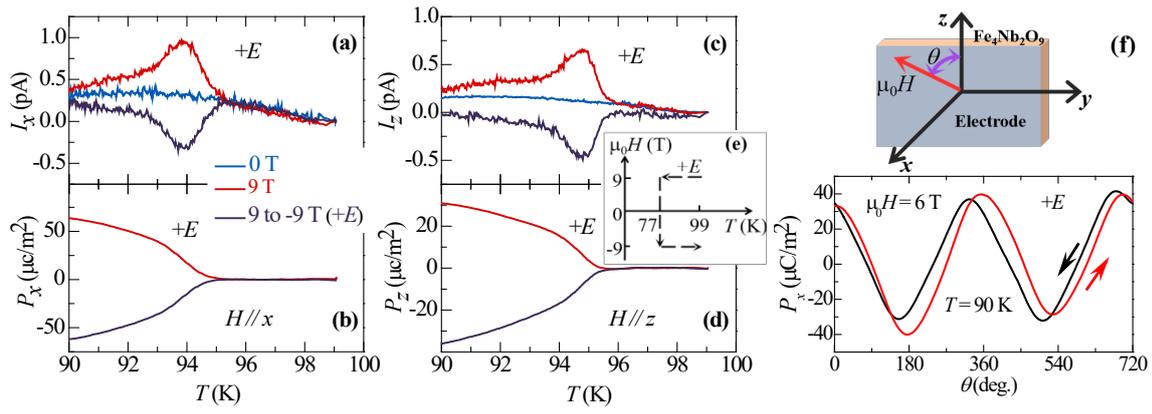

Figure 4 Jing Zhang *et al*.